\input phyzzx
\input maggieref
\newcount\mongocount
\mongocount=1
\def\Figure#1#2#3{
      \vbox to #3in{\hsize=#2in
        \vfil
         \includegraphics{#1}
    }
}
\def\figcap#1#2{
\vtop{\tenpoint\singlespace
\hsize=#1in\smallskip\noindent Figure\ \ \the\mongocount.\ \  #2
\global\advance\mongocount by 1\bigskip}}
\def\mongofigure#1#2#3#4#5{\centerline{\Figure{#1}{#2}{#3}
\figcap{#4}{#5}}}

\hoffset=0.375in
\overfullrule=0pt

\def\max{{\rm max}}

\def\min{{\rm min}}

\def\pc{{\rm pc}}
\def\kpc{{\rm kpc}}

\twelvepoint
\font\bigfont=cmr17
\centerline{\bigfont Hubble Deep Field Constraint on Baryonic Dark Matter}
\bigskip
\centerline{\bf Chris Flynn}
\centerline{Tuorla Observatory, Piikki\"o, FIN-21500, Finland}
\smallskip
\centerline{{\bf Andrew Gould}\foot{Alfred P.\ Sloan Foundation Fellow}}
\centerline{Dept of Astronomy, Ohio State University, Columbus, OH 43210}
\smallskip
\centerline{\bf John N.\ Bahcall}
\centerline{Institute For Advanced Study, Princeton, NJ 08540}
\bigskip 
\centerline{cflynn@astro.utu.fi, gould@payne.mps.ohio-state.edu, jnb@ias.edu} 
\bigskip
\centerline{\bf Abstract}
\singlespace

	We use a new technique to search for faint red stars in the Hubble 
Deep Field (HDF) imaged by the Wide Field Camera
(WFC2) on the {\it Hubble Space Telescope}.
We construct a densely sampled stellar light profile from 
a set of undersampled images of bright stars in a low-latitude field. 
Comparison of this stellar light profile to densely sampled profiles of 
individual objects in the HDF constructed from multiple undersampled dithers
allows us to distinguish unambiguously between stars and galaxies to $I=26.3$.
We find no stars with $V-I>1.8$ in the outer 90\% of the volume probed. 
This result places strong and general 
constraints on the $I$ band luminosity of the constituents of the 
Galactic dark halo:
$$M_I > 15.9 + {5\over 3}\log\biggl(f{0.5 M_\odot\over M}\biggr)
\qquad (V-I>1.8),$$
where $M$ is the mass of the objects and $f$ is their density as a fraction of
the local halo density, taken to be  
$\rho_0=9\times 10^{-3}\,M_\odot\,\pc^{-3}$.
If the halo is made of white dwarfs, this limit implies that these objects
have $M_V\gsim 18.4$ and $V-I\gsim 2.5$. That is, they are $\gsim 2$ magnitudes
fainter than the end of the disk white dwarf sequence. 
Faint red dwarfs account for $<1\%$ of the Galactic dark halo for $M_I<14$, 
and $<6\%$ for $M_I<15$ at the 95\% confidence level.  The density of
inter-galactic Local Group stars is at least a factor 3000 smaller than the
density of local Population II stars.

Subject Headings:  dark matter -- gravitational lensing -- stars: low mass,
brown dwarfs, luminosity function
\endpage

\chapter{Introduction}

	The detection of $\sim 7$ candidate microlensing events by the MACHO 
collaboration (Alcock et al.\ 1993, 1995, 1996a,b) during two years of 
observations toward the Large Magellanic Cloud (LMC) implies that a 
significant fraction $(0.2\lsim f \lsim 1)$ of the
dark halo of the Galaxy may be in the form of massive compact halo objects
(MACHOs).  
The mean observed time scale of
the events is $\VEV{t_e}\sim 37\,$days, where $t_e$ is the Einstein radius
crossing time.  This time scale is substantially
longer than would be expected if the MACHOs have sub-stellar masses, 
$M\lsim 0.1\, M_\odot$, and if their velocity and spatial distributions
were in accord with a standard halo model.  Within the context of such a
model, the best estimate for the mean mass is $\VEV{M}\sim 0.4\,M_\odot$,
with a range $0.1\lsim \VEV{M}/M_\odot\sim 1$ at the $3\,\sigma$ level.
The EROS collaboration (Aubourg et al.\ 1993, 1995; Ansari et al.\ 1996), 
has detected two candidate events from a somewhat smaller data set and
with a somewhat shorter mean time scale $\VEV{t_e}=26\,$days.  The EROS data
are consistent with the results inferred from the MACHO data.  In particular,
the EROS experiment was less sensitive than MACHO to the relatively long
events of the type that MACHO detected.  These long 
duration events contribute heavily
to both the high estimate of the total MACHO density and the high inferred
mean mass of the detected objects.

	If there is a substantial population of MACHOs, then it is possible
that some of these objects may be luminous enough and close enough to be
directly detected in optical light.  Red dwarfs were a plausible candidate
of this type.
Bahcall et al.\ (1994, hereafter Paper I) searched for such
objects in deep images taken by the Wide Field Camera (WFC2) on the
{\it Hubble Space Telescope (HST)} to a limiting magnitude of $I=25.3$.
From the null results of that search, they concluded that faint red 
dwarfs $(M_I<14)$ make up $<6\%$ of the Galactic halo.  With additional 
plausible assumptions, Graff \& Freese (1996) 
have derived even stronger constraints using the same observational results.

	The new results of MACHO (Alcock et al.\ 1996b)
may point in a different direction: white
dwarfs (WDs).  Although several arguments appear to constrain the WD 
contribution to the halo to be $<10\%$ (Charlot \& Silk 1995) or $<25\%$ 
(Adams \& Laughlin 1996), the fact that MACHO may be detecting objects in
the WD mass range means that this candidate must be taken seriously.
More generally, the nature of the detected objects, whatever they are, can
be probed or constrained by searching for intrinsically faint stars.

	The Hubble Deep Field (HDF) (Williams et al.\ 1996)
taken with WFC2 on {\it HST} provides a unique
window on the universe (Bahcall, Guhathakurta, \& Schneider 1990;
Abraham et al.\ 1996; Colley \& Rhoads 1996).
	The extreme depth of the HDF, which has an
equivalent exposure time $\sim 10$ greater than the field analyzed in
Paper I, provides an unprecedented opportunity to find faint stellar objects.
The principal advantage of going deep is that it allows one to search
for faint stars in regions of the color-magnitude diagram (CMD) which are
virtually devoid of the stars that populate the standard Galactic components.
The lack of ordinary disk and spheroid 
stars at very faint magnitudes is a result of the
finiteness of the Galaxy.  Thus, by restricting attention to objects within
1.67 mag of the magnitude limit, one probes 90\% of the available volume
for intrinsically faint stars while eliminating nearly all of the stars
from previously well-studied Galactic populations.
Of course, by going
deep one also increases the total volume probed (for candidate objects
of fixed luminosity) but this advantage is secondary: a survey of 16 
fields each 2 mag less deep would have the same volume and would require
only 40\% of the telescope time.  However, these fields would contain of
order 30 dwarf and sub-dwarf Galactic stars within 1.67 mag of the magnitude 
limit, where we have made the estimate based on actual star counts in the
{\it HST} Large Area Multi-Color Survey (``Groth Strip'', $l=96^\circ$, 
$b=60^\circ$).
Hence,
the HDF provides a truly unique opportunity to search for intrinsically
faint stars.

	In \S\ 2, we describe our technique for discriminating stars from
extended objects to a magnitude limit of $I=26.3$.  In \S\ 3, we discuss 
our selection criteria and report that 
no stars are detected. 
In \S\ 4, we show that the lack 
of detections means that if the Galactic halo is composed of WDs of mass
$M=0.5\,M_\odot$, these must have $M_I> 15.9$ (or $M_V\gsim 18.4)$.  
For red dwarfs and brown dwarfs of mass
$0.08\,M_\odot$, the limit is stronger, $M_I>17.2$.

\chapter{Finding Stars in the HDF}

	Objects can be detected in HDF down to $I\sim 28$.
The overwhelming majority of these very faint objects are galaxies.  Hence,
for purposes of studying galaxies, no star/galaxy discrimination is required.
For star counts however, it is essential to distinguish unambiguously between
the handful of stars being detected and the thousands of background
galaxies.  Because distinguishing stellar from extended profiles requires
$\sim 5$ times more photons than just detection, the magnitude limit
for the star counts will be much brighter than the detection limit, as we
quantify below.
 
   In order to classify objects on WFC2 images, we developed and tested in 
Paper I an effective procedure for separating stars and galaxies based on the 
radial profiles of the objects. The empirical stellar radial profile was
determined from a large number of stars that appear on a WFC2 exposure
at low galactic latitude.  Many stars (falling at random places
relative to the pixel grid) were required because the WFC2 poorly samples
the point spread function (PSF). 
(The full width half maximum of the PSF is approximately 1.2 pixels.
See Fig.\ 1, below.)\ \ 

   The exposures of the HDF were taken in groups at several different
positions on the sky (these groupings are termed ``dithers''), each
offset slightly relative to the HDF field center. This procedure
improves the smoothness of the flatfielding corrections and helps
improve the spatial resolution.  The HDF team released special
``drizzled'' images in four filters, F300, F450, F606, and F814.  These
images were produced using a ``drizzling'' technique (Fruchter \&
Hook 1996), which takes into account the shifts and rotations between
the individual dither positions and geometric distortions at the image
plane, while preserving the flux.
 
   Since drizzling involves the rebinning of undersampled data, the
radial profiles of sources in the drizzled image are slightly
different from those in the original images. Hence, to find the stars on
HDF we developed a procedure that combines the depth of the drizzled
image with the well-understood PSF properties of the raw data.
 
  We obtained the flatfielded and bias-subtracted frames from the Space
Telescope Science Institute.
The observing log indicates that some of the exposures were
affected by significant scattered light from Earth, which appears as a
characteristic X-pattern in the images. The HDF team ignored these
frames in constructing their first-release drizzled images, and we
followed their example in this respect. (By exposure time, 6\% of the
F606 and 25\% of the F814 frames suffer from significant scattered light). 
Because
of the pointing and tracking accuracy of HST, the images taken at each
dither position are precisely aligned. This alignment makes stacking and
cleaning (i.e.\ removing hot pixels and cosmic rays from) the images in each
dither straightforward, and we used the same procedure as described in Paper I
to create cleaned images at each dither position.  The total
exposure time in F814 was 92,200 seconds in 39 separate exposures,
taken at 8 separate dither positions.  In F606 the total exposure time
was 110,050 seconds in 105 separate exposures taken at 11 separate
dither positions.
  
  We searched the drizzled F814 band image for sources down to a
magnitude limit of $I=26.5$, 
beyond which the images on the dithered images become too noisy to classify
reliably.  We required that the ratio of peak to total flux be at least
half as large as the value found for known stars.  This initial selection
yielded 629 objects in all.  Each object was then
located on the 19 dither images (8 in F814 and 11 in F606). The radial
profile of the object was determined independently on each of the 19
dither images, and these were plotted against the empirical
stellar profile for the filters F814 and F606 (as determined in Paper I).

\FIG\starprof{
(a) A densely sampled light profile of a test object 
({\it circles}) is compared
to the template profile of a star ({\it solid line}) that is constructed
from a WFC2 image of stars in a low-latitude field (see Paper I).  
Each circle represents
one pixel in one of the 8 WFC2 dither images of HDF taken with the F814 filter.
The distance (in pixels) from the center of the star to
the center of the pixel is plotted against the fraction of all star light
from the dither image falling on that pixel.  Note that even though the
individual dither images are highly undersampled, the combined profile
is well sampled to well within 1 pixel.  The close agreement with the
template indicates that the object is stellar.  (b) Same as (a) except that 
only one of the eight dither images is shown.  
(c) Profile constructed from
the single drizzled image.  Vertical and horizontal axes have been adjusted
to account for the difference in pixel size.  Note that the profile is
nearly identical to the template profile except within 0.6 pixels of the
center.
}
\topinsert
\mongofigure{ps.fig1}{6.4}{6.5}{6.4}{
(a) A densely sampled light profile of a test object 
({\it circles}) is compared
to the template profile of a star ({\it solid line}) that is constructed
from a WFC2 image of stars in a low-latitude field (see Paper I).  
Each circle represents
one pixel in one of the 8 WFC2 dither images of HDF taken with the F814 filter.
The distance (in pixels) from the center of the star to
the center of the pixel is plotted against the fraction of all star light
from the dither image falling on that pixel.  Note that even though the
individual dither images are highly undersampled, the combined profile
is well sampled to well within 1 pixel.  The close agreement with the
template indicates that the object is stellar.  (b) Same as (a) except that 
only one of the eight dither images is shown.  
(c) Profile constructed from
the single drizzled image.  Vertical and horizontal axes have been adjusted
to account for the difference in pixel size.  Note that the profile is
nearly identical to the template profile except within 0.6 pixels of the
center.
}
\endinsert
	The key to star/galaxy separation is that we can locate 
the center of a stellar image to an accuracy of better than 0.1 pixels, using 
the symmetry of the stellar image about its center.  Thus, for any given star
and on any given dither, the radial profile of the image is sampled
in many neighboring pixels; the separations 
correspond to different distances between the stellar center and pixel 
centers.  We construct diagrams that give the light intensity of 
images in pixels whose centers  are located at different distances from the 
center of the stellar light.  We then superimpose the diagrams from the 
separate dithers and thereby create a densely sampled profile.

  Figure \starprof(a) shows an object of moderate signal to noise, which we
classify as a star (this object appears on Chip 4 of the drizzled
images at $(x,y)=(350,598)$ and has $I=23.71$, $V-I=1.32$).  The 
densely sampled radial
profile from the 8 independently centered dither images (shown as
circles) matches the empirical stellar profile (shown as a line) very
well.  The radial profile is well sampled because the center
of the object falls on the various dither images over a good range of
spatial positions relative to the pixel grid.  Figure \starprof(b) shows the
profile of this star from a single dither.  Note that for only three pixels 
does the pixel center fall within one pixel of the center of the star.
However, these three points are sufficient to distinguish stars from galaxies 
provided that the galaxies are extended by $\gsim 1$ pixels, i.e.
$\gsim 0.\hskip-2pt ''1$.  We used plots of this type in Paper I and in 
Gould, Bahcall, \& Flynn (1996) to identify stars in 22 WFC2 fields.
In Figure \starprof(c), we show the profile of the same
object as seen in the drizzled image, where we take into account the
different pixel size (drizzled pixels are a factor of 0.402
smaller).  The drizzled profile begins to deviate from the WFC2 profile
at about 0.6 pixels.

	We compared the densely sampled profile with the drizzled profile
for several dozen stellar and nearly stellar objects.  We found that
for objects $I\lsim 25.5$, stars could be easily distinguished from galaxies
using either profile.  However, for $I\gsim 25.5$ we found several objects
whose densely sampled profiles are clearly non-stellar, but which could
not be distinguished from stars using the drizzled profile.
  Because of the difficulties in classifying objects on the drizzled
image, we identify candidate objects on the drizzled image but do the
classification on the 19 dither images.
 
   Diagrams similar to Figure \starprof(a) were made for each of the 629 
objects found
on the drizzled F814 frame, and CF and AG independently classified
them as either ``star'', ``galaxy'', ``quasi-stellar'', or ``?''.  
The designation
``quasi-stellar'' means that the object is clearly not a star, but deviates 
from a stellar profile only within 1 pixel.  

Occasionally, one
or more of the 19 images could not be used for accidental reasons
(like a bad pixel) and those cases were handled separately. Photometry
was carried out after the classification, using the same aperture size
and calibration from F606 and F814 to $I$, $V-I$ as described in Paper I.
Comparison of the classifications by CF and AG showed that we could classify 
stars
and galaxies with confidence to $I=26.2$. We found a total of 17 stellar
objects to this limit.  In addition, we found 1 blue stellar object beyond
the magnitude limit ($I=26.5$, $V-I=0$).  All stellar objects satisfying
the $I$ band magnitude limit were easily detected and photometered in $V$.

	The photometric errors, as determined from the scatter of the
photometric measurements made from different dithers, is $\sim 0.08$ mag
at the magnitude limit and $\sim 0.03$ mag at $I=25$.  The uncertainties
induced by photometric errors are therefore several orders of magnitude
smaller than the Poisson uncertainties.

\chapter{Selection Criteria}

\FIG\cmd{
Color-magnitude diagram ($V-I$ vs $I$) of stellar objects detected in HDF.
The errors are not shown since they would be of order or smaller than the 
size of the points.
The selection criteria deliminating the ``halo zone'' are shown as a box at 
the lower right.  These
are $I<26.30$ (magnitude limit for star/galaxy discrimination), 
$I>24.83$ (eliminate ordinary Galactic stars while keeping 90\% of the
volume probed), and $V-I>1.8$ (color region of interest for dark halo
candidates).
}
\topinsert
\mongofigure{ps.cmd}{6.4}{5.5}{6.4}{
Color-magnitude diagram ($V-I$ vs $I$) of stellar objects detected in HDF.
The errors are not shown since they would be of order or smaller than the 
size of the points.
The selection criteria deliminating the ``halo zone'' are shown as a box at 
the lower right.  These
are $I<26.30$ (magnitude limit for star/galaxy discrimination), 
$I>24.83$ (eliminate ordinary Galactic stars while keeping 90\% of the
volume probed), and $V-I>1.8$ (color region of interest for dark halo
candidates).
}\endinsert

	Figure \cmd\ is a CMD of the stars detected
in the HDF together with the adopted selection criteria, which are described
below.

	First, we restrict attention to red stars, $V-I>1.8$, which includes
M dwarfs, and is expected to include brown dwarfs and old WDs.  
The local density of old {\it disk} WDs
has been measured by proper-motion studies to $M_V=17$ 
(corresponding to $V-I\sim 1.8$)
(Liebert, 
Dahn, 
\& Monet 
1988).  
Halo WDs, which are expected to be fainter than this limit, would not have
shown up in this study because their proper motions $\mu$ would generally 
exceed the selection limit of $\mu<2.\hskip-2pt '' 5\,\rm yr^{-1}$
(or perhaps somewhat smaller).

	Second, we set the magnitude limit at $I_\max=26.30$.  As discussed in
the previous section, we find that our star/galaxy separation is reliable
for $I<26.2$.  The brightest object above this limit that lies in the
adopted color range and is not obviously a galaxy has $I=23.39$.  We therefore
have detected all stars to $I=26.3$. 

	Third, we establish a lower magnitude limit $I_\min$ so as to exclude
ordinary Galactic stars.  As discussed in the introduction, the HDF is so
deep that one expects very few ordinary stars near the magnitude limit.
By setting $I_\min = I_\max - 1.67= 24.63$, we therefore exclude the regions
of the CMD that are heavily populated with previously well-studied stars, 
while preserving 90\% of the total volume.  
	Figure \cmd\ shows that, with the above-described criteria,
the color-magnitude diagram is devoid of stars in the region,
$ 24.63 < I < 26.30,$ $V-I>1.8$

	Finally, we note that for $V-I\lsim 0.7$, 
there are $\sim 60$ compact non-stellar objects that lie near the 
magnitude limit.  Since some of these objects deviate from point sources
only in the inner fraction of a pixel ($<0.\hskip-2pt''1$), one must wonder
whether there are not other objects of this class which appear perfectly
stellar.  Indeed, we find three such faint stellar objects with A-star 
colors and $I\sim 26$ (see Fig.\ \cmd), and are currently investigating 
the nature of these
extremely compact objects as a whole.  The presence of this class of object 
complicates the search for faint blue stars.  However, since they are
well blueward of the ``halo zone'', they have no impact on the principal
results of the present paper.

\chapter{Limits}

	Since we detect no objects with $V-I>1.8$, we can rule out at the 95\%
confidence level any model that predicts 3.0 or more detections in this 
regime.  In particular, since the HDF goes 1 magnitude deeper than the field
used in Paper I, we can immediately extend the results derived there.
For red dwarfs brighter than $M_I=15$ (the faintest red dwarf ever seen,
Monet et al.\ 1992) the halo fraction must be $f<6\%$.  For $M_I<14$, the
fraction is $\lsim 1\%$.

	To further interpret these results, consider a class of objects each 
with mass $M$ that comprise a fraction
$f$ of the dark halo, and so have a local number density 
$n = f\rho_0/M$ where $\rho_0=9\times 10^{-3}M_\odot\,\pc^{-3}$ 
is taken as the local halo density (Bahcall, Schmidt, \& Soneira 1983).  
Now suppose that these objects are detectable out to a
distance $d$ within the HDF with angular area $\Omega = 4.4\,\rm arcmin^2$.
Then the number of expected detections $N_{\rm exp}$ is
$$N_{\rm exp} = {1\over 3}n\Omega d^3 = 
3 f\biggl({M\over 0.5\,M_\odot}\biggr)^{-1}\biggl({d\over 1.1\,\kpc}\biggr)^3,
\eqn\nexp$$
where we have assumed initially that the halo density is uniform over the
region probed.  In fact, equation \nexp\ shows that one can begin to place
limits on objects that can be detected to $d\sim 1$--2 kpc, where the exact
distance limit $d$ depends only weakly on the details of the model.  
For simplicity, 
we adopt a halo with density $\propto r^{-2}$ which has a local density
$\rho_h = 0.86\rho_0$ at a distance of 1.5 kpc in the direction of the
HDF ($l=126^\circ$,$b=55^\circ$).  Since the magnitude limit is 
$I<I_\max=26.30$ and since 10\% of the volume is excluded by selecting 
$I>I_\min=24.63$, we find a limit
$$M_I > 15.9 + {5\over 3}\log\biggl(f{0.5 M_\odot\over M}\biggr)
\qquad (V-I>1.8).\eqn\extraequ $$
That is, for objects brighter than this limit there are more than 3 
expected detections, contrary to the observations.

	For WDs of mass $M=0.5\,M_\odot$ and $V-I>1.8$, these limits imply
$M_I> 15.9 + 1.67\log(f)$.  To interpret this limit as a limit on the
halo fraction, we assume that the observed linear color-magnitude relation 
for the red end of the disk WDs (Monet et al.\ 1992), 
$M_I = 12.9 + 1.2(V-I)$, 
can be extended to the fainter halo WDs.
We then find limits on the halo fraction $f$ of
$$f <  0.31 \times 10^{0.72[(V-I)-1.8]}.\eqn\flim$$ 
Hence, under these assumptions, a full WD standard halo is ruled out for
$V-I<2.5$ ($M_V<18.4$) and a 33\% WD standard halo is ruled out for $V-I=1.8$
$(M_V=16.9)$, both at the 95\% confidence level.

	We now compare the sensitivity of our results with the measurements 
of and constraints on luminous halo objects obtained with a variety of 
techniques as reviewed by Mould (1996).  Generally, if one is interested 
in objects that are at least as bright as the end of the locally observed
sub-dwarf sequence ($M_V\sim 15$, $V-I\sim 3$), then there are several
probes that are at least as 
sensitive than the one presented here.  For example,
Dahn et al.\ (1995) use proper-motion selected stars to construct a luminosity
function.  Based on this model, we expect to find $\lsim 1$ star with 
$V-I > 1.8$.  Similarly, the deep color-selected ground-based survey of 
Boeshaar
et al.\ (1994) can detect such objects over $\sim 3$ times the volume probed
by HDF (although such surveys are subject to significant contamination by
faint galaxies).  The real strength of the HDF is its sensitivity to
intrinsically faint objects, near the limit \extraequ. These would have  
avoided detection in proper-motion surveys.  For red objects ($V-I\gsim 3.5$),
the volume probed by HDF is larger than for any ground-based photometric
surveys and, of course, HDF is free of galaxy contamination.

	Finally, we note that the HDF star counts can be used to constrain
the density of Local Group stars. 
Local Group giants and sub-giants 
$M_I\lsim 1.5$, $0.6\lsim V-I\lsim 1.5$ could be seen to a distance 
$d\sim 0.9\,$Mpc.  In the outer 90\% of this volume, there are no more than
2 such stars observed, implying that their density must be $<7\times 10^{-11}\,
\pc^{-3}$ at the 95\% confidence level.  Comparing this limit to the density
of local spheroid giants $\sim 2\times 10^{-7}\,\pc^{-3}$ 
(Morrison, 1993; Flynn \& Fuchs 1994) we constrain the ratio of the densities 
of Local Group to galactic spheroid stars to be $<1/3000$, about an order of
magnitude lower than the limits obtained by Richstone et al.\ (1992).

{\bf Acknowledgements}:  
We are grateful to K.\ Griest and P.\ Sackett for valuable comments.  
We thank the Director of STScI, Robert Williams, and the HDF-team
for skillful execution of the HDF concept and for rapidly reducing
and making available the superb images.
A.~G.\ was supported in part by NSF grant AST 9420746.
J.\ N.\ B.\ was supported in part by NASA grant NAG5-1618.
The work is based in large part
on observations with the NASA/ESA Hubble Space Telescope, obtained
at the Space Telescope Science Institute, which is operated by the
Association of Universities for Research in Astronomy, Inc. (AURA), under
NASA contract NAS5-26555.

\endpage
\Ref\Abr{Abraham, R.\ G., Tanvir, N.\ R., Santiago, B.\ X., Ellis, R.\ S., 
\& Glazebrook, K.\ 1996, MNRAS, in press}
\Ref\adams{Adams, F.\ C.\ \& Laughlin, G.\ 1996, ApJ, submitted}  
\Ref\Al{Alcock, C., et al.\ 1993, Nature, 365, 621}
\Ref\Alcock{Alcock, C., et al.\ 1995, Phys.\ Rev. Lett.\ 74, 2867}
\Ref\Alcock{Alcock, C., et al.\ 1996a, submitted}
\Ref\Alcock{Alcock, C., et al.\ 1996b, in preparation}
\Ref\Aubourg{Aubourg, E., et al.\ 1993, Nature, 365, 623}
\Ref\Aubourg{Aubourg, E., et al.\ 1995, A\&A, 301, 1}
\Ref\Ansari{Ansari, R.\ et al.\ 1996, A\&A, in press}
\Ref\BFG{Bahcall, J.N., Flynn, C., Gould, A., \& Kirhakos, S.\ 1994, ApJ, 
435, L51 (Paper I)}
\Ref\BGS{Bahcall, J.\ N., Guhathakurta, P.\ \& Schneider, D.\ P.\ 1990,
Science, 248, 178}
\Ref\BSS{Bahcall, J.\ N., Schmidt, M., \& Soneira, R.\ M.\ 1983, ApJ,
265, 730}
\Ref\boe{Boeshaar, P.\ C., Tyson, T., \& Bernstein, G.\ 1994, BAAS, 185, 1346}
\Ref\cs{Charlot, S.\ \& Silk, J.\ 1995, ApJ, 445, 124}
\Ref\cr{Colley, W.\ N.\ \& Rhoads, J.\ E.\ 1996, preprint}
\Ref\dah{Dahn, C.~C., Liebert, J.\ W., Harris, H., \& Guetter, H.\ C.\ 1995, 
p.\ 239, An ESO Workshop on: the Bottom of the Main Sequence and Beyond,
C. G. Tinney ed. (Heidelberg: Springer)}
\Ref\ff{Flynn, C.\ \& Fuchs, B.\ 1994, MNRAS, 270, 471}
\Ref\www{Fruchter, A.\ S.\ \& Hook, 1996
http://www.stsci.edu/ftp/observer/hdf/combination/drizzle.html}
\Ref\gbm{Gould, A., Bahcall, J.\ N., \& Flynn, C.\ 1996, ApJ, 465, 000}
\Ref\GF{Graff, D., \& Freese, K.\ 1996, ApJ, 456, L49}
\Ref\ldm{Liebert, J., Dahn, C.\ C., \& Monet, D.\ G.\ 1988, ApJ, 332, 891}
\Ref\mon{Monet, D.\ G., Dahn, C.\ C., Vrba, F.\ J., Harris, H.\ C.,
Pier, J.\ R., Luginbuhl, C.\ B., \& Ables, H.\ D.\ 1992, AJ, 103, 638}
\Ref\mor{Morrison, H.\ 1993, AJ, 106, 587}
\Ref\mould{Mould, J.\ 1996, PASP, 108, 35}
\Ref\rich{Richstone, D.\ O., Gould, A., Guhathakurta, P., \& Flynn, C.\
1992, ApJ, 388, 354}
\Ref\wil{Williams, R.\ et al.\ 1996, Science with the Hubble Space Telescope 
II, P.\ Benvenuti, F.\ D.\ Macchetto, \& 
E.\ J.\ Schreier, eds., in press  (Baltimore: STScI)}

\refout
\endpage
\end